\def\Y{\Upsilon}
\def\B{{\cal B}}
\def\bb{b\bar b}
\long\def\simplex#1#2#3#4{
\begin{figure}[#1]
   \begin{center}
   \quad\\[-1cm]
   \quad
   \hbox{
   \quad 
   \parbox[t]{\hsize}{ \psfig{figure=#2,width=10.0cm} 
   \caption[]{\small  \label{fig:#3} #4 }}
   }
   \quad
   \end{center} 
\end{figure}
}
\long\def\ssimplex#1#2#3#4#5{
\begin{figure}[#1]
   \begin{center}
   \quad\\[-1cm]
   \quad
   \hbox{
   \quad 
   \parbox[t]{\hsize}{ \psfig{figure=#2,width=#5} 
   \caption[]{\small  \label{fig:#3} #4 }}
   }
   \quad
   \end{center} 
\end{figure}
}
\long\def\duplex#1#2#3#4#5#6#7{
\begin{figure}[#1]
   \begin{center}
   \quad\\[-1cm]
   \quad
   \hbox{
   \quad 
   \parbox[t]{7.5cm}{ \psfig{figure=#2,width=8.0cm} 
   \caption[]{\small  \label{fig:#3} #4 }}
   \quad
   \parbox[t]{7.5cm}{ \psfig{figure=#5,width=8.0cm} 
   \caption[]{\small \label{fig:#6} #7 }}
   }
   \quad
   \end{center} 
\end{figure}
}
\long\def\quadruplex#1#2#3#4#5#6#7#8#9{
\begin{figure}[#1]
   \begin{center}
   \quad\\[-2cm]
   \quad
   \hbox{
   \quad 
   \parbox[t]{7.5cm}{ \psfig{figure=#2,width=8.0cm} 
   \caption[]{\small  \label{fig:\1} #3 }}
   \quad
   \parbox[t]{7.5cm}{ \psfig{figure=#4,width=8.0cm} 
   \caption[]{\small \label{fig:\2} #5 }}
   }
   \\[-1cm]
   \hbox{
   \quad 
   \parbox[t]{7.5cm}{ \psfig{figure=#6,width=8.0cm} 
   \caption[]{\small  \label{fig:\3} #7 }}
   \quad
   \parbox[t]{7.5cm}{ \psfig{figure=#8,width=8.0cm} 
   \caption[]{\small \label{fig:\4} #9 }}
   }
   \quad
   \end{center} 
\end{figure}
}

\documentclass[aps,prd,preprint,superscriptaddress,tightenlines]{revtex4}
\usepackage{graphicx}% Include figure files
\usepackage{dcolumn}% Align table columns on decimal point
\usepackage{bm}% bold math
\usepackage{epsfig}

\begin{document}

\preprint{CLEO CONF 02-06}
\preprint{ICHEP02 ABS948}

\title{First Observation of $\Y(1D)$ States}
\thanks{Submitted to the 31$^{\rm st}$ International Conference on High Energy
Physics, July 2002, Amsterdam}

%-------- INSERT HERE ------------
% Your author list goes here  REMOVE EVERYTHING to END INSERT and
% replace with your authorlist (ask cleoac).

\author{S.~E.~Csorna}
\author{I.~Danko}
\affiliation{Vanderbilt University, Nashville, Tennessee 37235}                             
\author{G.~Bonvicini}
\author{D.~Cinabro}
\author{M.~Dubrovin}
\author{S.~McGee}
\affiliation{Wayne State University, Detroit, Michigan 48202}                               
\author{A.~Bornheim}
\author{E.~Lipeles}
\author{S.~P.~Pappas}
\author{A.~Shapiro}
\author{W.~M.~Sun}
\author{A.~J.~Weinstein}
\affiliation{California Institute of Technology, Pasadena, California 91125}                
% ...
\collaboration{CLEO Collaboration}
\noaffiliation

%-------- END INSERT ------------

%please hard code the date when you have a final draft and submit to CLEOAC
\date{July 19, 2002}

\begin{abstract} 
% Insert abstract here.
The CLEO III experiment has recently accumulated
a large statistics sample of $4.73\cdot 10^6$ $\Y(3S)$ decays.
We present the first evidence for the production of the triplet
$\Y(1D)$ states in the four-photon cascade,
$\Y(3S)\to\gamma\chi_b(2P)$, 
$\chi_b(2P)\to\gamma\Y(1D)$, 
$\Y(1D)\to\gamma\chi_b(1P)$, 
$\chi_b(1P)\to\gamma\Y(1S)$,
followed by the $\Y(1S)$ annihilation
to $e^+e^-$ or $\mu^+\mu^-$.
The signal has a significance of  $9.7$ standard deviations.
The measured product branching ratio for these five decays,
$(3.3\pm0.6\pm0.5)\cdot 10^{-5}$, is consistent with the
theoretical estimates.
We see a $6.8$ standard deviation signal for a state with a mass
of $10162.2\pm1.6$~MeV/$c^2$, consistent with the $\Y(1D_2)$ assignment.

We also present improved measurements of the
$\Y(3S)\to\pi^0\pi^0\Y(1S)$ branching ratio
and the associated di-pion mass distribution.
\end{abstract}

\maketitle

\section{Introduction}

Long-lived $b\bar b$ states are especially well suited for
testing QCD via lattice calculations \cite{LatticeQCD}
and effective theories
of strong interactions, like potential models \cite{PotentialModels}.
The narrow triplet $S$ states, $\Y(1S)$, $\Y(2S)$ and
$\Y(3S)$, were discovered in 1977 in proton-nucleus
collisions at Fermilab \cite{UpsilonDiscovery}. 
Later, they were better resolved and studied at various
$e^+e^-$ storage rings.  
Six triplet $P$ states, $\chi_b(2P_J)$ and $\chi_b(1P_J)$ with
$J=0,1,2$, were discovered in radiative
decays of the $\Y(3S)$ and $\Y(2S)$ 
in 1982 \cite{chib2pdiscovery} 
and  1983 \cite{chib1pdiscovery}, respectively.
There has been no observations of new narrow $\bb$ states since
then despite the large number of such states predicted below the
open flavor threshold (see Fig.~\ref{fig:level}).

In this paper, we present the first observation of the 
$\Y(1D)$ states.
Except for the yet unobserved $\Y(2D)$ states, these are the only
long-lived $L=2$ mesons in nature.
They are produced in a two-photon
cascade starting from the $\Y(3S)$ resonance:
$\Y(3S)\to\gamma\chi_b(2P_J)$, 
$\chi_b(2P_J)\to\gamma\Y(1D)$. 
To suppress photon backgrounds from $\pi^0$s, which are copiously
produced in gluonic annihilation of the $b\bar b$ states,
we select events with two more subsequent photon transitions,
$\Y(1D)\to\gamma\chi_b(1P_J)$, 
$\chi_b(1P_J)\to\gamma\Y(1S)$, followed
by the $\Y(1S)$ annihilation to either 
$e^+e^-$ or $\mu^+\mu^-$
(see Fig.~\ref{fig:level}).
The product branching ratio for these five decays in sequence was
predicted by Godfrey and Rosner \cite{GodfreyRosner}
to be $3.76\cdot 10^{-5}$.

\ssimplex{htbp}{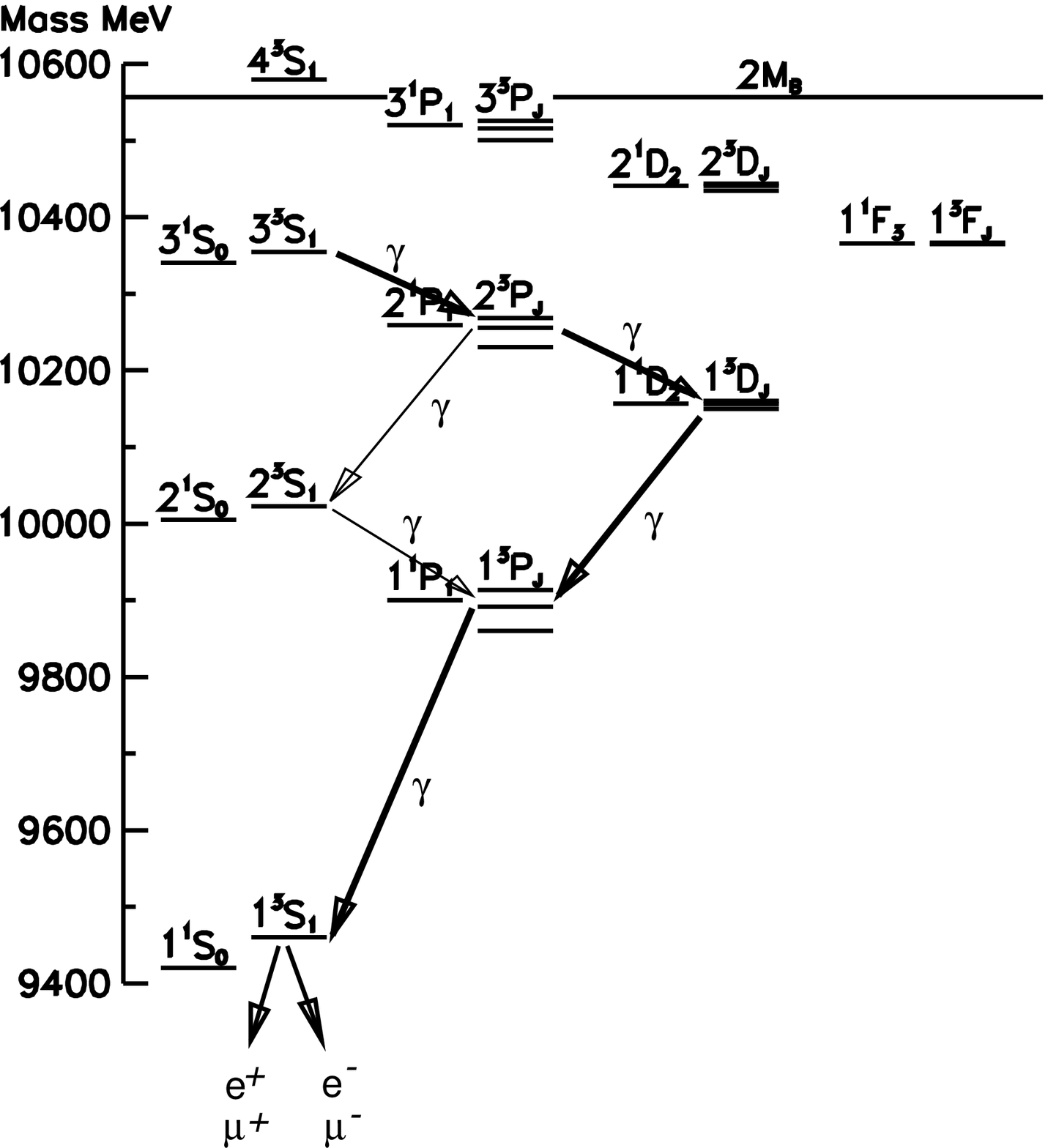}{level}{
$\bb$ mass levels as predicted by one of the potential models.
The levels are denoted by their spectroscopic labels, 
$n^{2S+1}L_J$, where $n$ is the radial quantum number ($n=1, 2, \dots$),
$S$ is the total quark spin ($S=0$ spin singlets, $S=1$ spin triplets),
$L$ is the orbital angular momentum ($L=S, P, D, \dots$) and $J$ is the total
angular momentum of the state ($\vec{J}=\vec{S}+\vec{L}$).
The four-photon transition sequence via the $\Y(1D)$ states is shown.
An alternative route in four-photon cascade via the $\Y(2S)$ state
is also displayed. 
}{10cm}

The CLEO III experiment has recently accumulated $4.73\cdot 10^6$
$\Y(3S)$ decays, which makes the detection of such rare
processes possible.
The CLEO III sample constitutes roughly a ten-fold increase in $\Y(3S)$
statistics compared to the CLEO II data set \cite{CLEOII3sExcl}, 
and roughly a four-fold increase compared to the integrated CUSB 
data set \cite{cusbp0p0}. 
Thanks to the good granularity and the large solid angle of the CLEO CsI(Tl) 
calorimeter, the CLEO III detection efficiency 
for these final states is about 4.5 times larger than in 
the CUSB detector.
Even though the CLEO III calorimeter is essentially the same
as in the CLEO II detector \cite{CLEOIIdetector}, 
the photon selection efficiency and
detector resolution were improved in the endcaps and in part of the
barrel calorimeter by the 
installation of a new, lower-mass, tracking system \cite{CLEOIIIDR}. 
Another change was the replacement of the time-of-flight system by a RICH
detector in the barrel part. 
Finally, the calorimeter endcaps were 
restacked and moved farther away from the interaction point to
accommodate a new, higher luminosity, interaction point optics.

\section{Data selection}

We select events with exactly four photons and two oppositely
charged leptons. The leptons must have momenta of at least 3.75 GeV/$c$.
We distinguish between electrons and muons by their 
energy deposition in the calorimeter. Electrons must have a high
ratio of energy observed in the calorimeter to the momentum measured
in the tracking system ($E/p>0.7$).
Muons are identified as minimum ionizing particles, and required to
leave $150-550$ MeV of energy in the calorimeter.
Stricter muon identification does not reduce background
      in the final sample, since all significant background
      sources contain muons.
Each photon must have at least 60 MeV of energy. We also ignore
all photons below 180 MeV in the calorimeter region closest to the
beam, because of the spurious photons generated by 
beam-related backgrounds.
The total momentum of all photons and leptons in each event
must be balanced to within 300 MeV/$c$.
The invariant mass of the two leptons must be consistent 
with the $\Y(1S)$ mass within $\pm300$ MeV/$c^2$.
Much better identification of the $\Y(1S)$ resonance is
obtained by measuring  the mass of the system recoiling
against the four photons. The average resolution of the recoil mass is
17 MeV/$c^2$. Cuts on the recoil mass are described below.
The mass resolution of the produced $\Y(1D)$ state
depends on the measurement of the energies of the two lowest energy photons
in the event. Thus, we require that at least one of them is detected in the barrel 
part of the calorimeter, where the energy resolution is best.
The energy resolution as determined by the fit to the $\Y(3S)\to\gamma\chi_b(2P_J)$
photon lines is  $\sigma_{E_\gamma}=(4.6\pm0.2)$ MeV 
for $E_\gamma=100$ MeV \cite{2gammaICHEP}.

\section{$\Y(3S)\to\pi^0\pi^0\Y(1S)$ transitions}

The most prominent signal among
$\Y(3S)\to\gamma\gamma\gamma\gamma\Y(1S)$
decays are $\Y(3S)\to\pi^0\pi^0\Y(1S)$
transitions. To identify these decays, we try all three
combinations of four photons into two-photon pairs.
We find the minimal $\pi^0$ mass deviation chi-squared
among these combinations:
$\chi^2_{\pi^0\pi^0}= min\,\left\{
[(M_{\gamma\gamma}^{ij}-M_{\pi^0})/\sigma_{M_{\gamma\gamma}^{ij}}]^2+
[(M_{\gamma\gamma}^{kl}-M_{\pi^0})/\sigma_{M_{\gamma\gamma}^{kl}}]^2\right\}$.
After the requirement $\chi^2_{\pi^0\pi^0}<6$, the deviation of the four-photon
recoil mass from the $\Y(1S)$ mass, $\Delta M=M_{recoil\, 4\gamma} - M_{\Y(1S)}$,
is found. The distribution of the ratio, $\Delta M/\sigma(\Delta M)$, where
$\sigma(\Delta M)$ is the expected recoil mass resolution, 
is plotted in Fig.~\ref{fig:mdrpi0pi0}.
As can be seen from the sidebands of this distribution,
backgrounds from non-$\Y(3S)\to\Y(1S)$
transitions (e.g.\ radiative Bhabha and $\mu-$pair events) are small.
We subtract this background by fitting the recoil mass 
distribution with the signal shape extracted from the Monte Carlo
simulations. The fit is also shown in Fig.~\ref{fig:mdrpi0pi0}.
We observe  $737\pm28$ events with an efficiency of 13.6\%.
The efficiency was determined from the Monte Carlo simulations.
This sample contains a small fraction, 0.9\%, of photon cascade background
that we estimate from the cross-efficiency determined with the
Monte Carlo (below 1\%) and the branching ratios measured from the
data, as described below. After the background subtraction, we
obtain the following result for the product branching ratio:
$\B(\Y(3S)\to\pi^0\pi^0\Y(1S))
\cdot\B(\Y(1S)\to l^+l^-)=(5.67\pm0.22\pm0.35)\cdot10^{-4}$.
The first error is statistical and the second is systematic.
The systematic error is dominated by uncertainties in the Monte Carlo
simulation of the selection efficiency. It was determined by
varying the selection criteria.
Here, $\B(\Y(1S)\to l^+l^-)$ means either $\B(\Y(1S)\to e^+e^-)$
or $\B(\Y(1S)\to \mu^+\mu^-)$, since these two branching
ratios should be equal from lepton universality.
Using the world average value \cite{PDG}, $\B(\Y(1S)\to l^+l^-)=(2.43\pm0.06)\%$,
we can unfold our result for
$\B(\Y(3S)\to\pi^0\pi^0\Y(1S))=(2.33\pm0.09\pm0.16)\%$.
This result is consistent with, but more precise, than previous
measurements by CLEO II \cite{c2p0p0}, $(2.03\pm0.28\pm0.19)\%$, and
CUSB \cite{cusbp0p0}, $(2.3\pm0.4\pm0.3)\%$.
The efficiency-corrected invariant mass distribution of 
the two-pion system is shown in Fig.~\ref{fig:mpi0pi0}.
As in the previous measurements, some dynamical structure
is observed in this distribution.

\duplex{htbp}{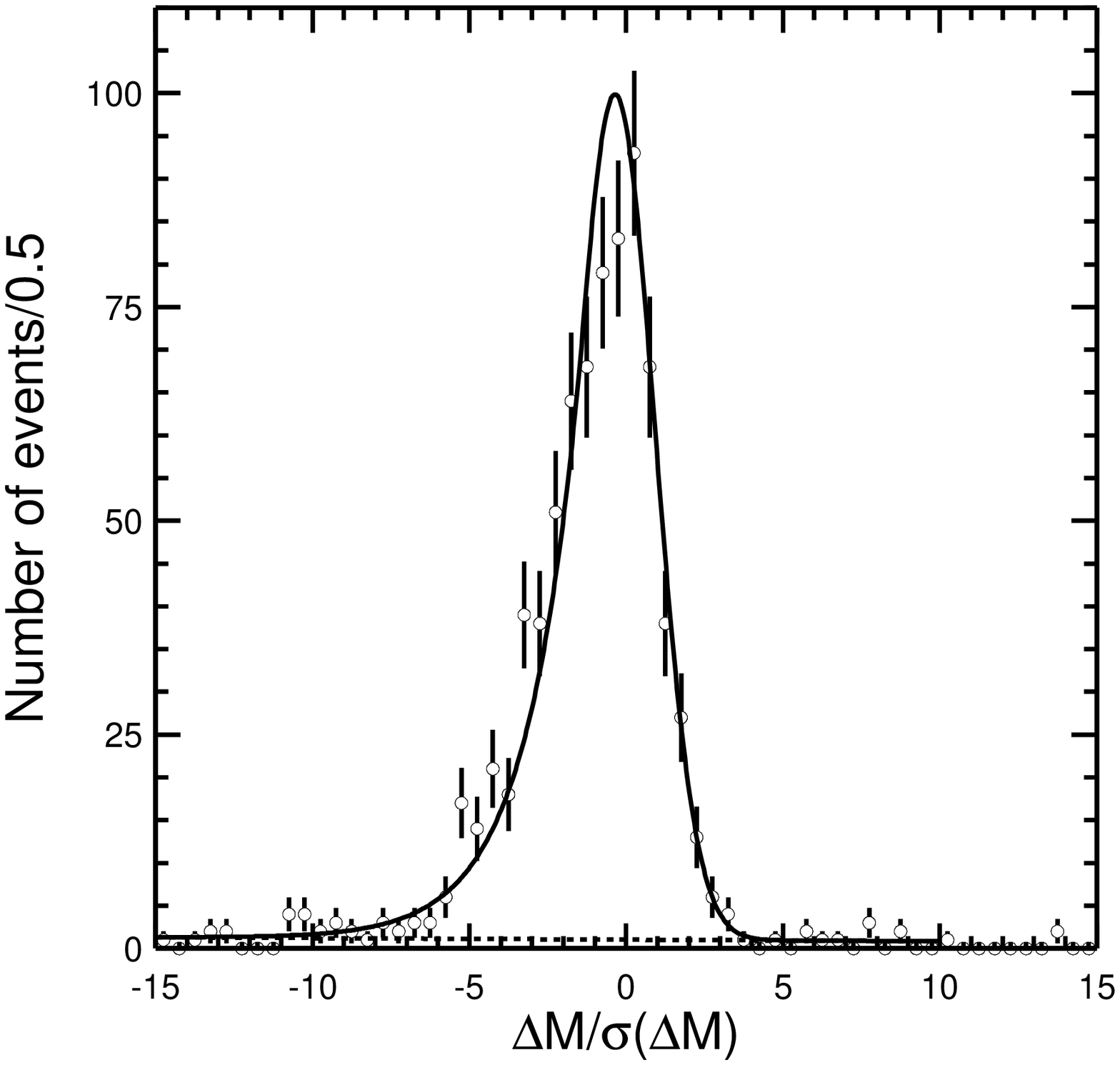}{mdrpi0pi0}{Distribution of 
the recoil mass deviation from the $\Y(1S)$ mass
for the $\Y(3S)\to\pi^0\pi^0\Y(1S)$ candidate events.
The solid-line represents the fit of 
a signal peak on top of the linear background (dashed-line).
}{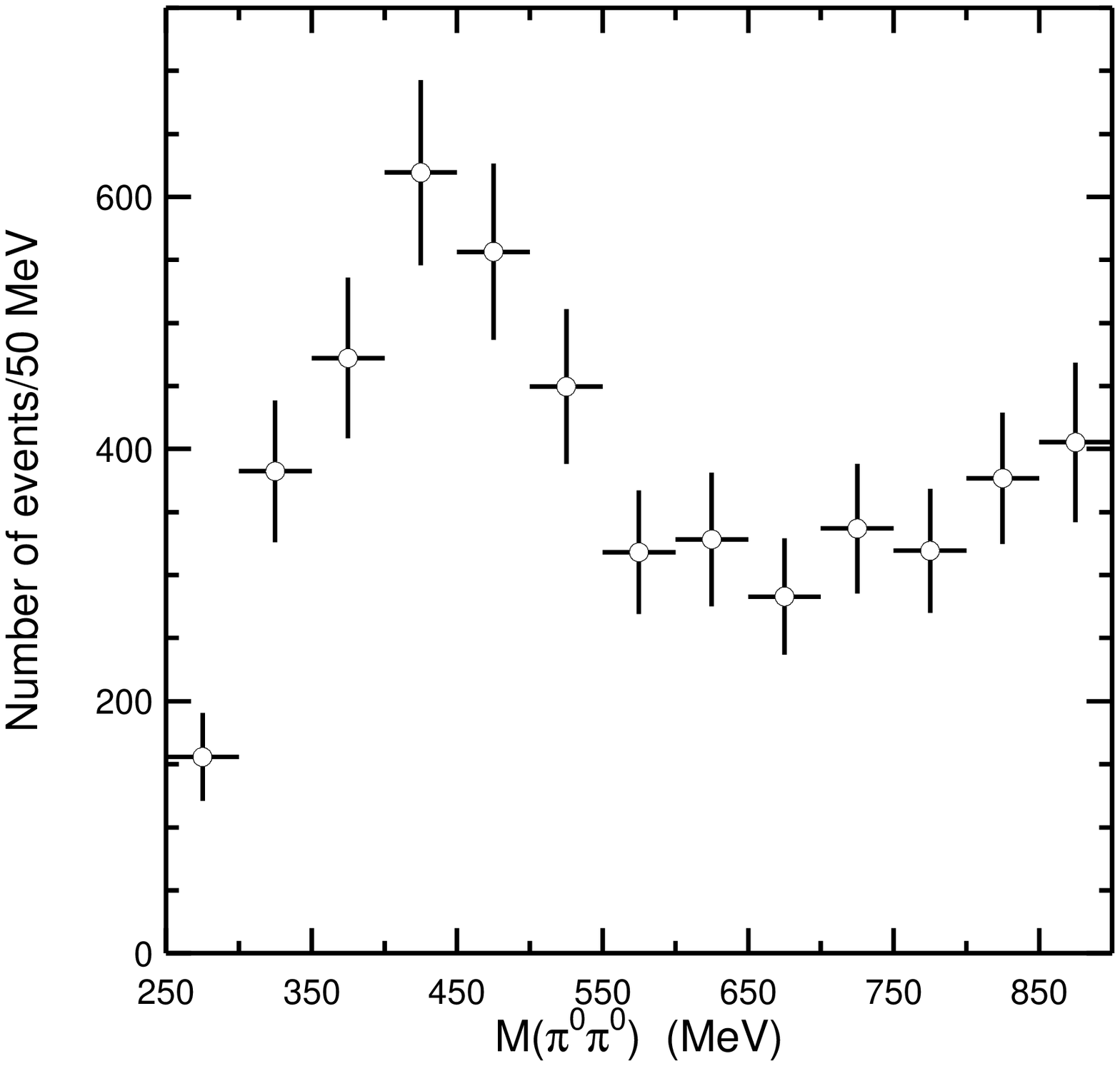}{mpi0pi0}{Efficiency-corrected distribution
of the $\pi^0\pi^0$ invariant mass
observed in the data.}

The determination of $\B(\Y(3S)\to\pi^0\pi^0\Y(1S))$ presented above
serves as a good cross-check of our experimental techniques.
In a separate paper submitted to this conference \cite{2gammaICHEP},
we also present the analysis of the two-photon cascades between the $\Y(3S)$
and the $\Y(1S)$ or $\Y(2S)$. Branching ratios for
two-photon cascades via $\chi_b(2P)$ and $\chi_b(1P)$ states 
were determined with improved precision compared to  previous
experiments. These data, together with the 
$\pi^0\pi^0$ transitions, were used to
obtain the calibration of absolute photon energies in the $60-800$ MeV
range. Using the well known masses of the $\Y(1S)$, $\Y(2S)$ and
$\Y(3S)$ \cite{PDG}, together with the recoil mass distributions like
the one shown in Fig.~\ref{fig:mdrpi0pi0}, we calibrated
the photon energies to a precision of $0.35\%$.
The photon energies observed in the two-photon cascade data
for $\Y(3S)\to\gamma\chi_b(2P_J)$ agreed well with the
previous measurements.
Thus, the systematic effects in both rate and photon energy
measurements are well under control.

\section{Evidence for the $\Y(1D)$ states}

To look for four-photon cascades between the $\Y(3S)$ and
$\Y(1S)$, we must suppress backgrounds from the $\pi^0\pi^0$ 
transitions. We first require, 
$\chi^2_{\pi^0\pi^0}>6$. 
In addition, every photon pair must have an invariant mass at least
two standard deviations away from the nominal $\pi^0$ mass.

To look for $\Y(1D)$ events, we constrain events to be
consistent with a photon cascade from the $\Y(3S)$ to 
the $\Y(1S)$ via one of the $\chi_b(2P_J)$ and
one of the $\chi_b(1P_J)$ states ($J=0, 1, 2$).
The masses of the $P$ states are well known from other 
measurements \cite{PDG}.
For each $J_{2P}$, $J_{1P}$ combination we calculate a chi-squared:
$$
\chi^2_{1D,\,J_{2P},\, J_{1P}}(M_{\Y(1D)})=\sum_{j=1}^4 
\left( \frac{E_{\gamma\,j}-E_{\gamma\,j}^{expected}(M_{\Y(1D)},\,J_{2P},\, J_{1P})}{
             \sigma_{E_{\gamma\,j}} } \right)^2,
$$
where $E_{\gamma\,j}$ are the measured photon energies;
$E_{\gamma\,j}^{expected}$ are the expected photon energies from
the masses of the $b\bar b$ states and the measured photon directions
in the cascade. The masses of the $\Y(1D)$ states are 
not know. Therefore, we minimize the above chi-squared with respect
to $M_{\Y(1D)}$ which is allowed to vary for each event.
The above formalism requires that we know how to order the four
photons in the cascade. While the highest energy photon must be due to the
fourth transition, and the second highest energy photon must be
due to the third transition, there is sometimes an ambiguity
in the assignment of the two lowest energy photons to the
first two transitions, since the range of photon energies
in the $\Y(3S)\to\gamma\chi_b(2P_J)$ decay overlaps the possible photon
energy range in the $\chi_b(2P_J)\to\gamma\Y(1D)$
transition. We choose the combination that minimizes 
the above chi-squared.
There are nine possible combinations of  $J_{2P}$, $J_{1P}$
spins. We try all of them and choose the one that produces
the smallest chi-squared, 
$\chi^2_{1D}=min\,\chi^2_{1D,\,J_{2P},\, J_{1P}}$.

In addition to the four-photon cascade via the $\Y(1D)$ states, our
data contain events with the four-photon cascade via the $\Y(2S)$ state:
$\Y(3S)\to\gamma\chi_b(2P_J)$, 
$\chi_b(2P_J)\to\gamma\Y(2S)$, 
$\Y(2S)\to\gamma\chi_b(1P_J)$, 
$\chi_b(1P_J)\to\gamma\Y(1S)$,
$\Y(1S)\to l^+l^-$
(see Fig.~\ref{fig:level}).
This is a coincidence of the two-photon cascades,
$\Y(3S)\to\gamma\gamma\Y(2S)$ and
$\Y(2S)\to\gamma\gamma\Y(1S)$, 
which were experimentally observed in the two-photon
cascades detected in the $\Y(3S)$ \cite{2gammaICHEP}
and $\Y(2S)$ data \cite{2scascades}.
The product branching ratio for this entire decay sequence
(including $\Y(1S)\to l^+l^-$)
is predicted by Godfrey and Rosner \cite{GodfreyRosner}
to be $3.84\cdot10^{-5}$, thus
comparable to the predicted $\Y(1D)$ production rate.
Combining the measured rates for 
$\B(\Y(3S)\to\gamma\gamma\Y(2S))\cdot\B(\Y(2S)\to l^+l^-)$ 
\cite{2gammaICHEP} and 
$\B(\Y(2S)\to\gamma\gamma\Y(1S))\cdot\B(\Y(1S)\to l^+l^-)$ 
\cite{2scascades} with the measurement of
$\B(\Y(2S)\to\mu^+\mu^-)$ \cite{PDG},
we obtain 
a factor $1.9\pm0.3$ higher value.
In these events, the second highest energy photon
is due to the second photon transition
(see Fig.~\ref{fig:level}).
Unfortunately, these events can be sometimes confused
with the $\Y(1D)$ events due to the limited 
experimental energy resolution. The second and
third photon transitions in the $\Y(2S)$ cascade sequence
can be mistaken for the third and second transition
in the $\Y(1D)$ cascade sequence.
Therefore, it is important to suppress the $\Y(2S)$ cascades.
We achieve this by finding the $J_{2P}$, $J_{1P}$ combination that 
minimizes the associated chi-squared for the $\Y(2S)$ hypothesis,
$\chi^2_{2S}=min\,\chi^2_{2S,\,J_{2P},\, J_{1P}}$,
where $\chi^2_{2S}$ is exactly analogous to $\chi^2_{1D}$ with
the $M_{\Y(1D)}$ replaced with $M_{\Y(2S)}$. 
We then require $\chi^2_{2S}>12$.
Notice that the masses of all intermediate states are known for
the $\Y(2S)$ cascade, thus this variable is more
constraining than $\chi^2_{1D}$. 
To further suppress the $\Y(2S)$ cascade  events, we construct a
quasi-chi-squared variable, $\chi^{2+}_{2S}$, that sums in quadrature
only positive deviations of the measured photon energies
from their expected values. This variable is less sensitive to 
fluctuations in the longitudinal and transverse energy leakage 
in photon showers that sometimes produce large negative energy deviations
and correspondingly a large $\chi^2_{2S}$ value.
With the additional cut $\chi^{2+}_{2S}>3$, the cross-efficiency
for $\Y(2S)$ events is reduced to 0.7\%\ and is further suppressed
by a factor of two when a cut of $\chi^2_{1D}<10$ is required.

\duplex{hbtp}{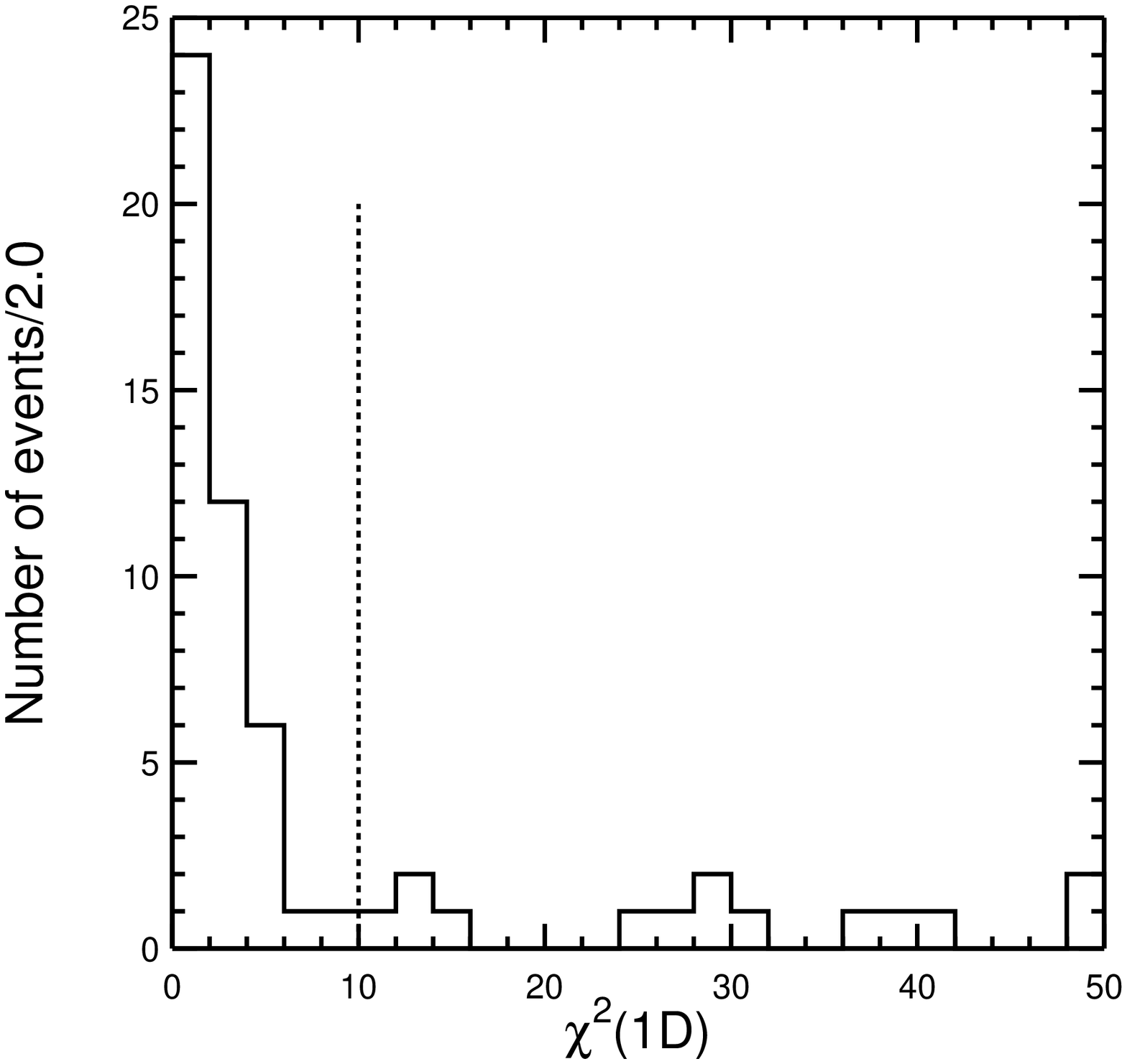}{chiddata}{
$\chi^2_{1D}$ distribution for the data.
The cut value used for the mass analysis is indicated by the 
vertical bar.}{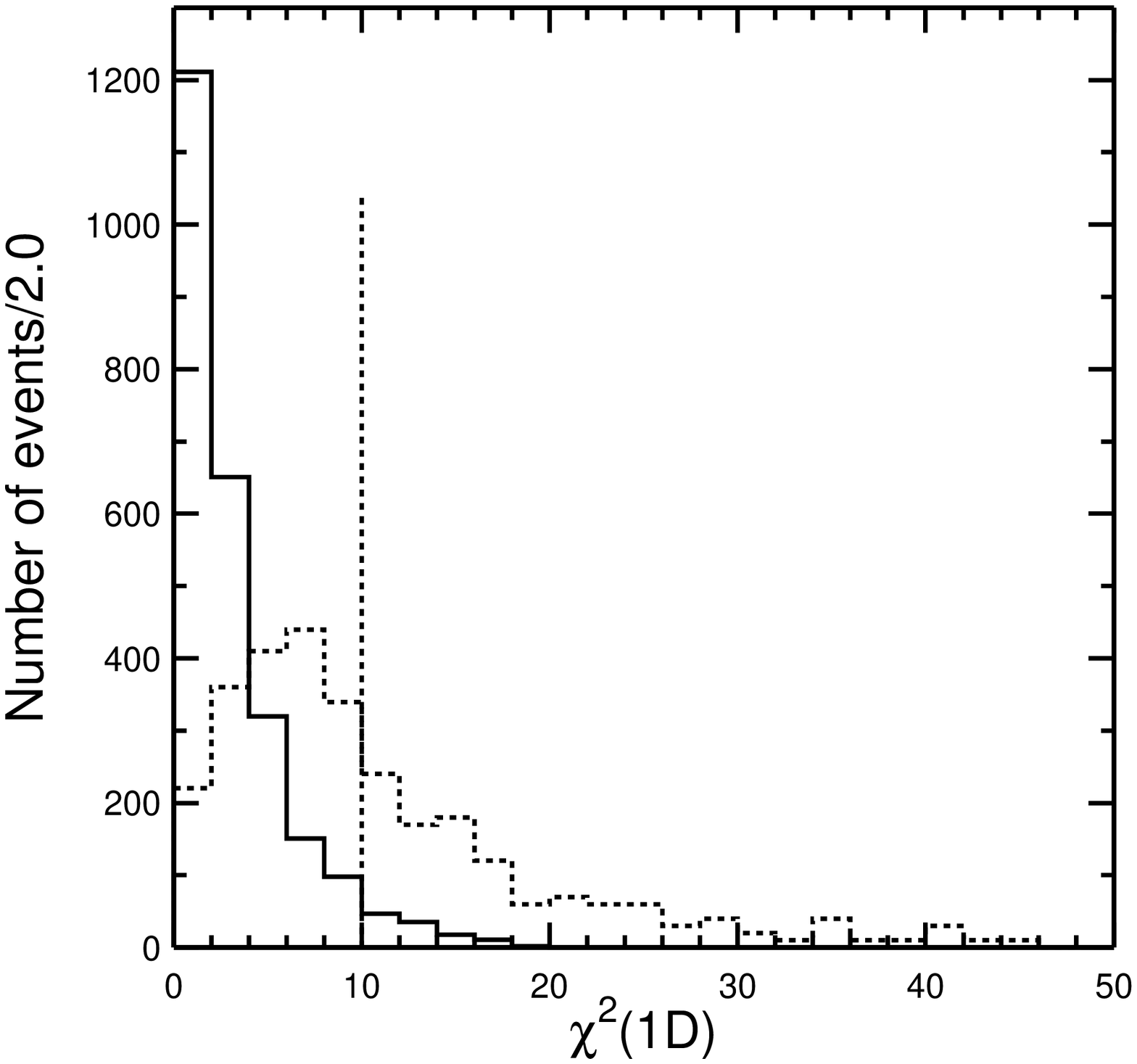}{chidmc}{
$\chi^2_{1D}$ distribution for the 
$\Y(1D)$ (solid-histogram) and $\Y(2S)$ cascade Monte Carlo (dashed-histogram). 
The latter distribution was scaled up by a factor
of 10 to be visible on the scale of the signal Monte Carlo.}

The data $\chi^2_{1D}$ distribution, 
after we require the four-photon recoil mass to be between
$-4$ and $+3$ standard deviations from the nominal $\Y(1S)$
mass, is shown in 
Fig.~\ref{fig:chiddata}. 
A narrow peak near zero is observed,
just as expected for $\Y(1D)$ events.
The signal Monte Carlo distribution for $\Y(1D)$ events is shown 
in Fig.~\ref{fig:chidmc}. 
The background Monte Carlo distribution for the $\Y(2S)$ cascades, 
after a factor of 10 enhancement relative to the 
$\Y(1D)$ normalization, is also shown for comparison.
Clearly, the $\Y(2S)$ cascade background cannot produce
as narrow a peak as observed in the data. 
The estimate of the number of $\Y(2S)$ cascade events passing 
the $\chi^2_{1D}<10$ requirement is $1.6-3.0$ events (depending
on the assumed $\Y(2S)$ product branching ratio).
We observe 44 events after this cut in the data.
The cross-efficiency for $\pi^0\pi^0$ events is $0.02\%$ 
in this sample, which gives a negligible contribution
to the signal region. 
A flat tail in the $\chi^2_{1D}$ distribution for the data, not observed
for the signal Monte Carlo, indicates that there is still some
background present from other processes like
radiative QED events. Since such processes can satisfy the multiple
mass constraints built into the $\chi^2_{1D}$ variable only by
a random coincidence, the distribution of that background varies slowly
in $\chi^2_{1D}$
except for the  bias towards lower values introduced
by the $\chi^2_{1D}$ minimization with respect to
$M_{\Y(1D)}$, $J_{2P}$ and $J_{1P}$.
We use the $\Y(3S)\to\pi^0\pi^0\Y(1S)$ Monte Carlo (with the $\pi^0$ veto cuts removed)
as a model of the $\chi^2_{1D}$ distribution for this background
component (see Fig.~\ref{fig:chidp0p0fit}).
Normalization of this background is fixed from the observed tail of the
$\chi^2_{1D}$ distribution.
This shape is likely to overestimate the amount of bias
in $\chi^2_{1D}$ towards the lower values since, unlike
the $\Y(3S)\to\pi^0\pi^0\Y(1S)$ process, the QED backgrounds
do not automatically satisfy the $M_{\Y(3S)}-M_{\Y(1S)}$ mass constraint
built into the chi-squared variable.
Therefore, we also used a linear background model, with the line parameters fixed
by a fit to the tail of the $\chi^2_{1D}$ distribution.
Depending on the background shape model, the $\chi^2_{1D}<10$ sample
has a $9.5-13.9\%$ background rate (including the $\Y(2S)$ cascade background).
A fit of the $\Y(1D)$ signal, on top of the backgrounds,
is shown in Fig.~\ref{fig:chidfit}.
The $\Y(2S)$ background contribution was fixed from 
the Monte Carlo simulations.
Normalization of the other backgrounds was left as a free parameter in the fit.
The fit gives a signal amplitude of $40.7\pm6.8$ events.
If we try to fit the data with the background contribution alone, 
the likelihood value of the fit drops dramatically. From the change
of the likelihood we estimate that the signal has a significance of
 9.7 standard deviations. Significant signals are observed in the 
$\gamma\gamma\gamma\gamma e^+e^-$ and  
$\gamma\gamma\gamma\gamma\mu^+\mu^-$ subsamples taken separately
($5.2\sigma$ and $8.2\sigma$, respectively).

\duplex{hbtp}{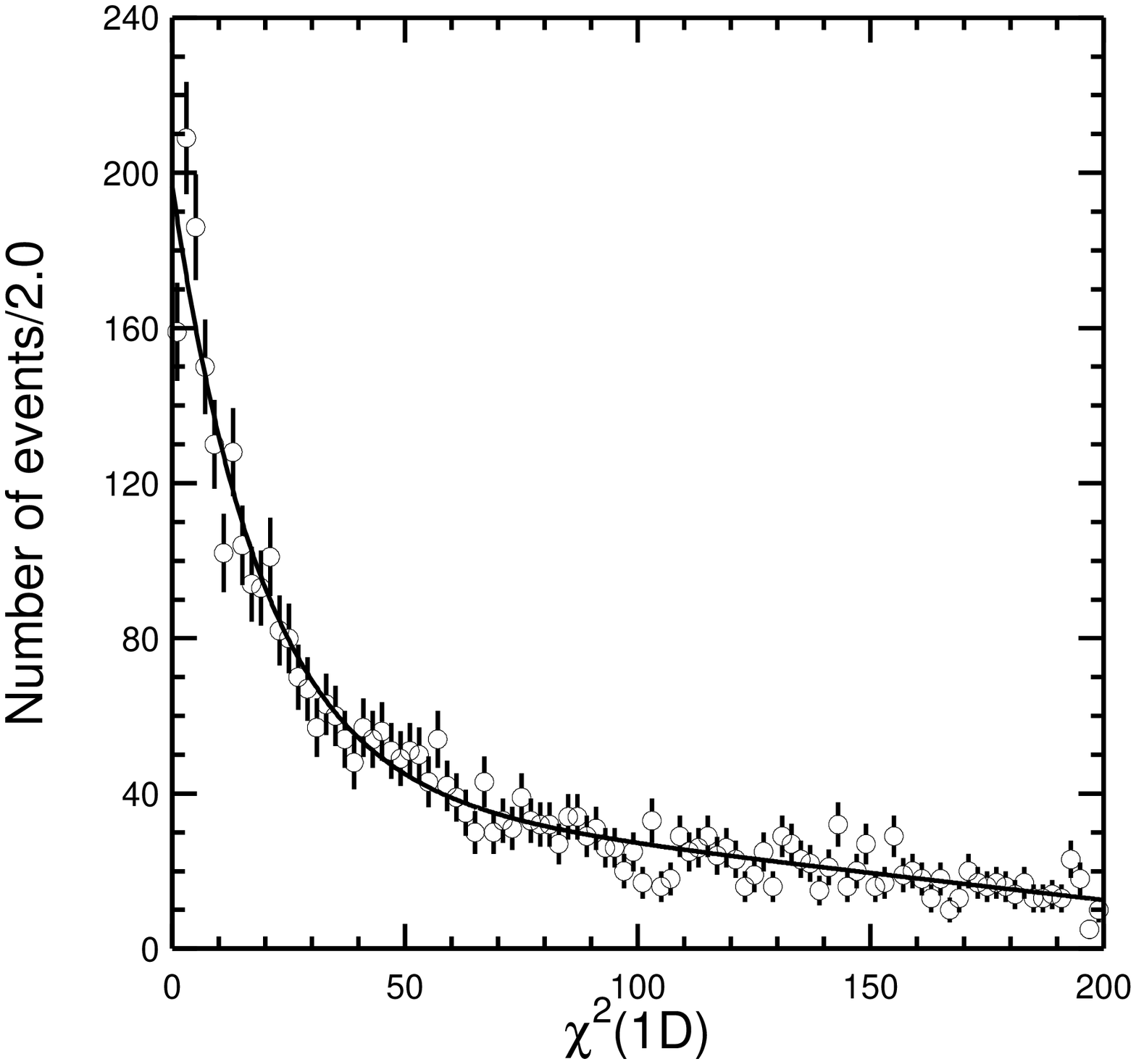}{chidp0p0fit}{Fit of a 
smooth curve to the $\chi^2_{1D}$ distribution observed
for the $\Y(3S)\to\pi^0\pi^0\Y(1S)$ Monte Carlo, with the
the $\pi^0$ veto cuts removed. 
}{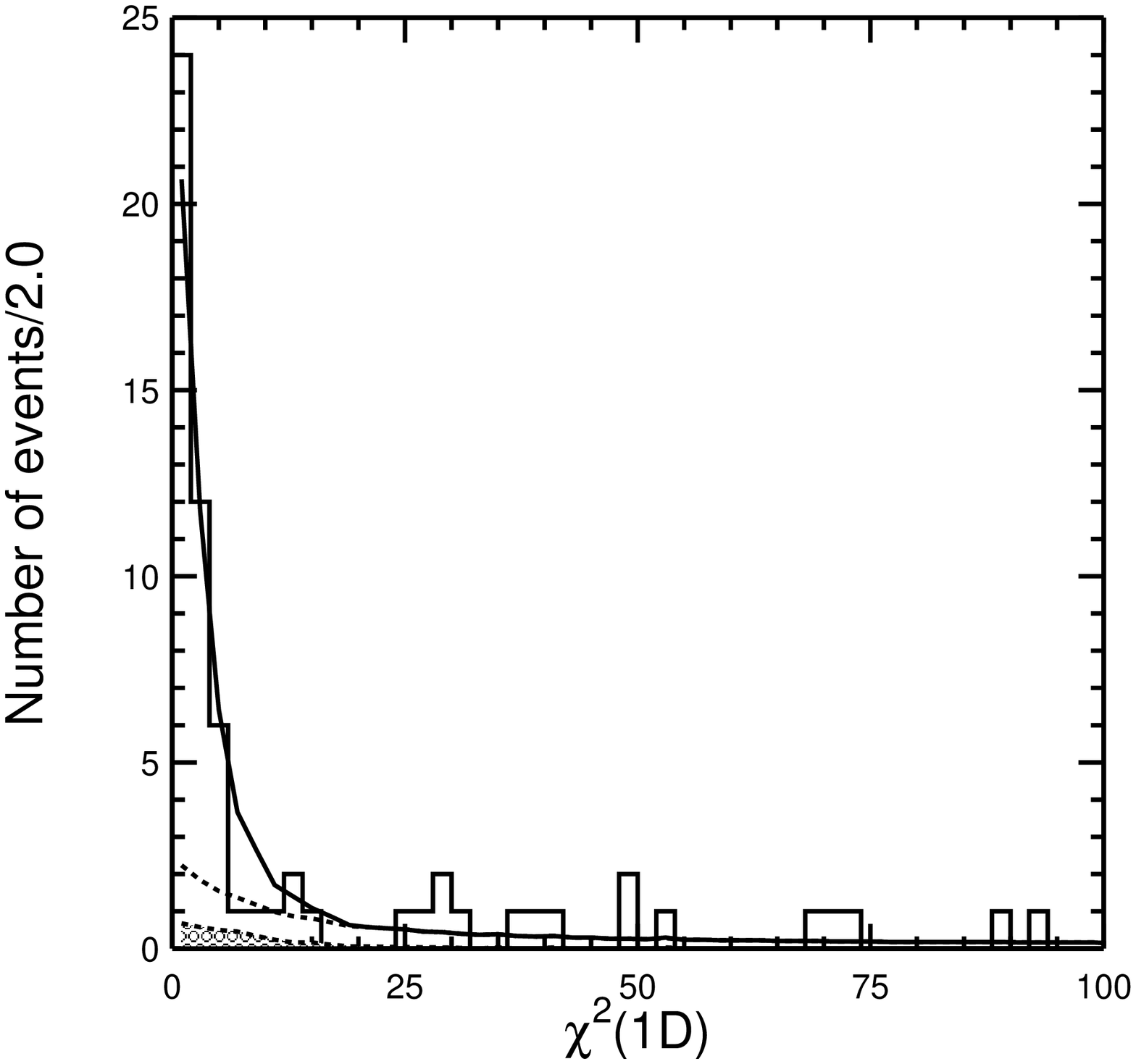}{chidfit}{Fit to the $\chi^2_{1D}$ distribution
in the data.
The solid-line represents the fit of the signal contribution on top
of the backgrounds.
The dashed-line represents the total background contribution.
The dashed-shaded area shows the $\Y(2S)$ cascade background.
}

\section{Mass analysis}

\simplex{htbp}{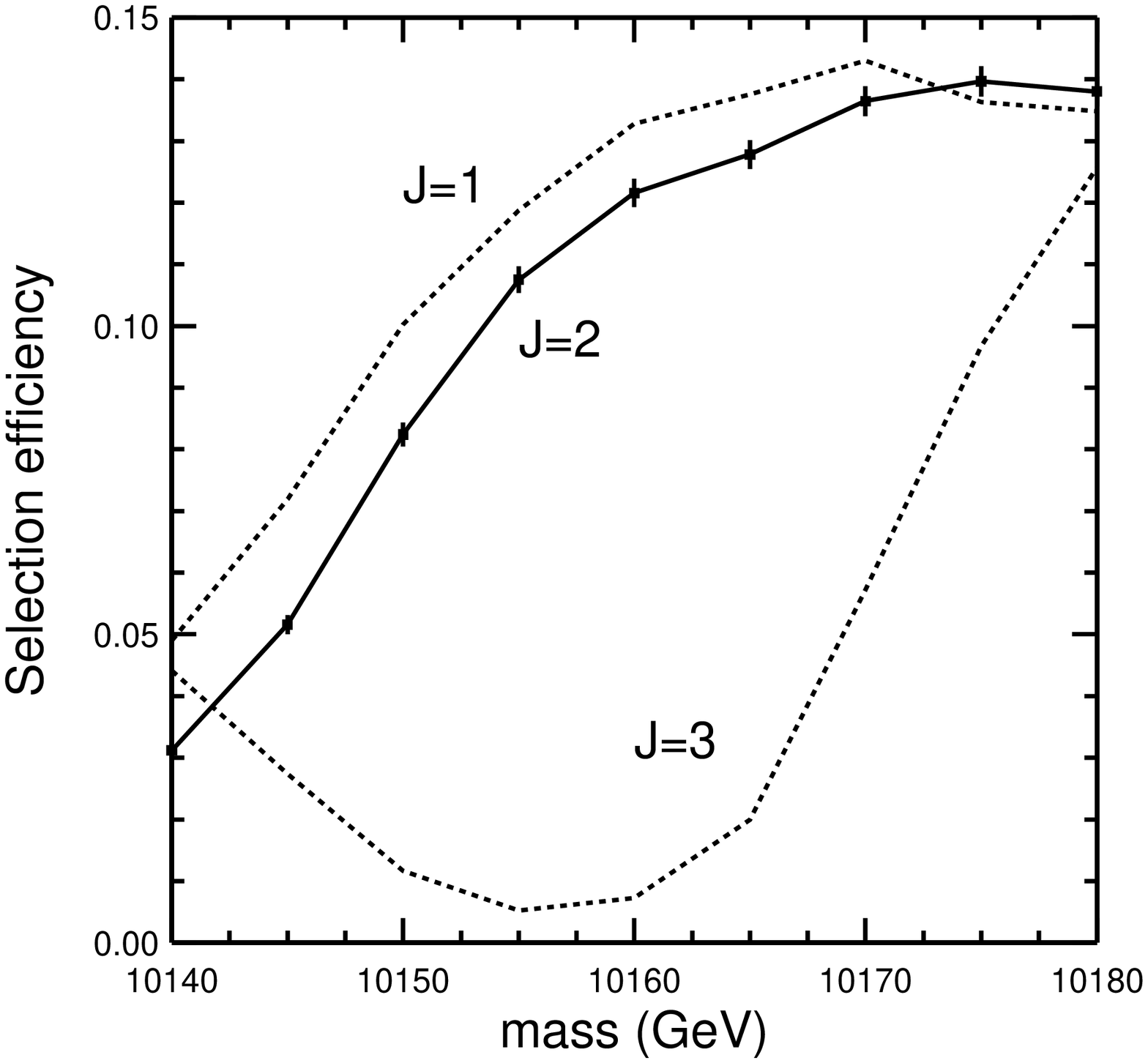}{effvsmass}{
Selection efficiency (with the $\chi^2_{1D}<10$ cut included) 
for $\Y(1D)$ states 
of different $J_{1D}$
as a function of the mass of the state.}

To turn the observed signal yield into a signal branching ratio,
we need to estimate the signal efficiency. Even though
the signal shape in the $\chi^2_{1D}$ variable is fairly independent
of the masses and spins of the produced $\Y(1D)$ states, the selection
efficiency does depend on these parameters via the cuts 
used to suppress the $\Y(2S)$ cascade background.
This is illustrated in Fig.~\ref{fig:effvsmass}, where we plot
the selection efficiency as a function of the mass of the
$\Y(1D)$ state. Three different curves are presented,
for $J_{1D}=1, 2$ and $3$. 
The $J_{1D}=2$ state is predicted by Godfrey and Rosner 
\cite{GodfreyRosner} 
to be produced with the highest rate, $2.6\cdot 10^{-5}$.
The $J_{2P}=1$, $J_{1P}=1$ combination is predicted to be the main
production and decay path for this state.
The selection efficiency for the $J_{1D}=2$ state
gradually increases with its mass, 
but is reasonably high in the entire search window.
The $J_{1D}=3$ state, predicted with a $0.8\cdot 10^{-5}$ rate,
can only be produced via the $J_{2P}=2$, $J_{1P}=2$ combination due to 
angular momentum conservation.
This happens to be an unfortunate combination, since for
a wide range of masses such a production fakes the $\Y(2S)$ cascade,
and the detection efficiency after the $\Y(2S)$ suppression
cuts is very low. We become sensitive to this state only
if its mass exceeds about \hbox{10170~MeV/$c^2$}.
Finally, the $J_{1D}=1$ state is predicted 
by Godfrey and Rosner 
to have the lowest product branching ratio, $0.4\cdot10^{-5}$.
This state is again predicted to be produced mostly via
the $J_{2P}=1$, $J_{1P}=1$ combination, and therefore our
efficiency for this state is similar to the $J_{1D}=2$ efficiency.

A straightforward way to measure the mass of the produced $\Y(1D)$ state 
is to calculate the mass of the system recoiling against the two lowest
energy photons in the event.
The distribution of the difference between the 
center-of-mass energy and this recoil mass, 
$E_{CM}-M_{recoil\,\gamma_1\gamma_2}$,
which should peak at $M_{\Y(3S)}-M_{\Y(1D)}$, is shown for
the Monte Carlo simulation of a $\Y(1D_2)$ state 
with a mass of 10160 MeV/$c^2$ in Fig.~\ref{fig:m3s1dmc}. 
This distribution can be well described with a Gaussian that has
a power-law tail on the low side (a so-called Crystal Ball Line Shape),
as illustrated in Fig.~\ref{fig:m3s1dmc}.
The asymmetric tail is
caused by transverse and longitudinal shower energy leakage
in the calorimeter.
The Gaussian part has a resolution of $\sigma=(6.8\pm0.3)$ MeV/$c^2$.

The mass value that minimizes $\chi^2_{1D}$ is also an estimate of the
true $\Y(1D)$ mass. The distribution of this mass for the
same Monte Carlo sample as above is shown in
Fig.~\ref{fig:m1dfmc}. A narrow peak at the correct mass is
observed with a resolution of $\sigma=(3.1\pm0.1)$ MeV/$c^2$.
The mass resolution is improved in this method due to 
the constraints from the well measured masses of the $P$ states
built into the $\chi^2_{1D}$ variable.
Unfortunately, there are also 
two satellite peaks, one on each side of the main peak.
These false peaks originate from photon 
energy fluctuations which
can make a wrong $J_{2P}$, $J_{1P}$ combination produce the
smallest chi-squared value. 

\global\def\1{m3s1dmc}
\global\def\2{m1dfmc}
\global\def\3{m3s1ddata}
\global\def\4{m1dfdata}
\quadruplex{htbp}{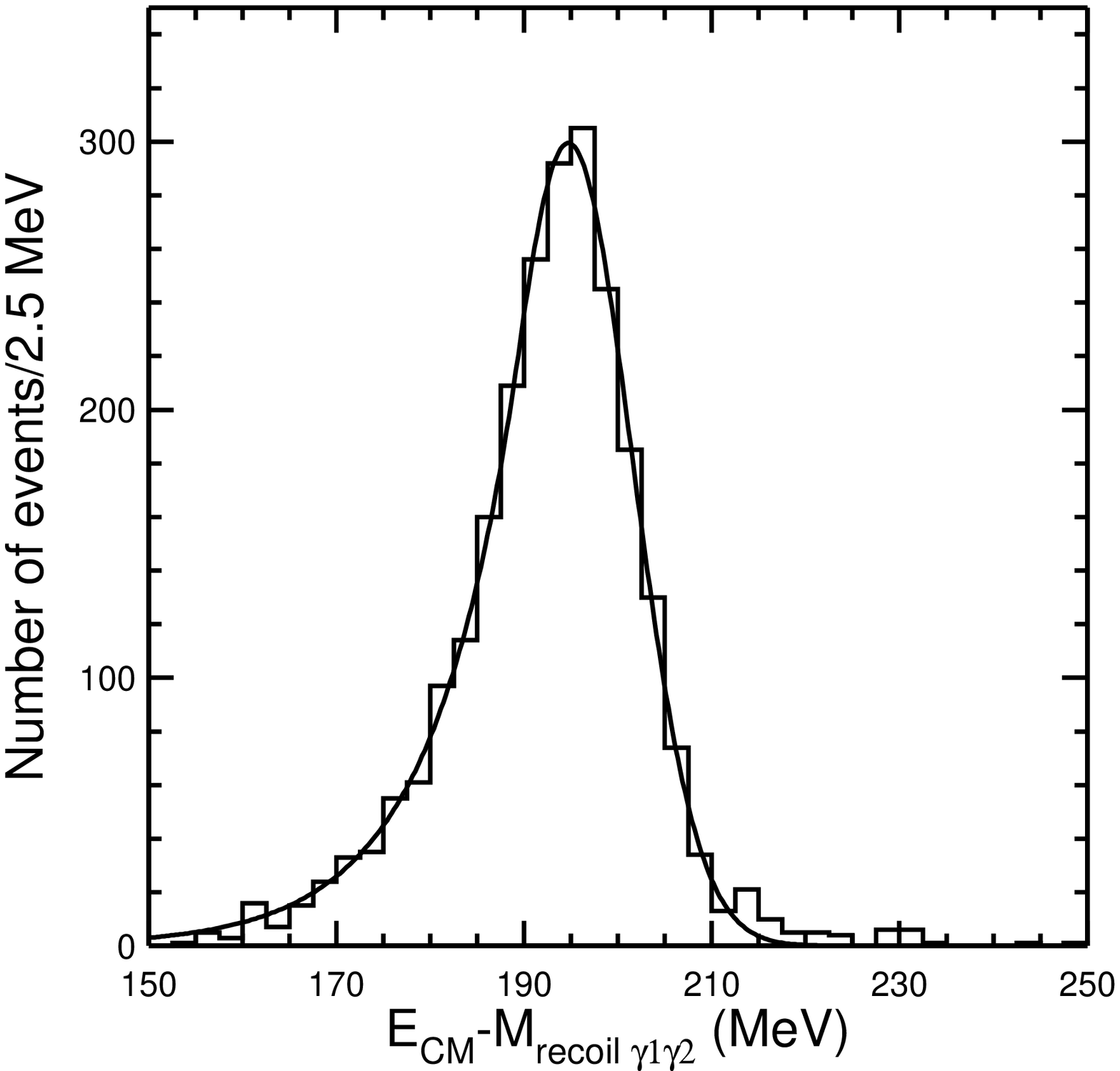}{
Distribution of the difference between the 
center-of-mass energy and the recoil mass
against the two lowest energy photons
for the $\Y(1D_2)$ Monte Carlo events generated with a
mass of 10160 MeV/$c^2$.
The solid-line illustrates the fit
of the Crystal Ball Line Shape to the distribution.
}{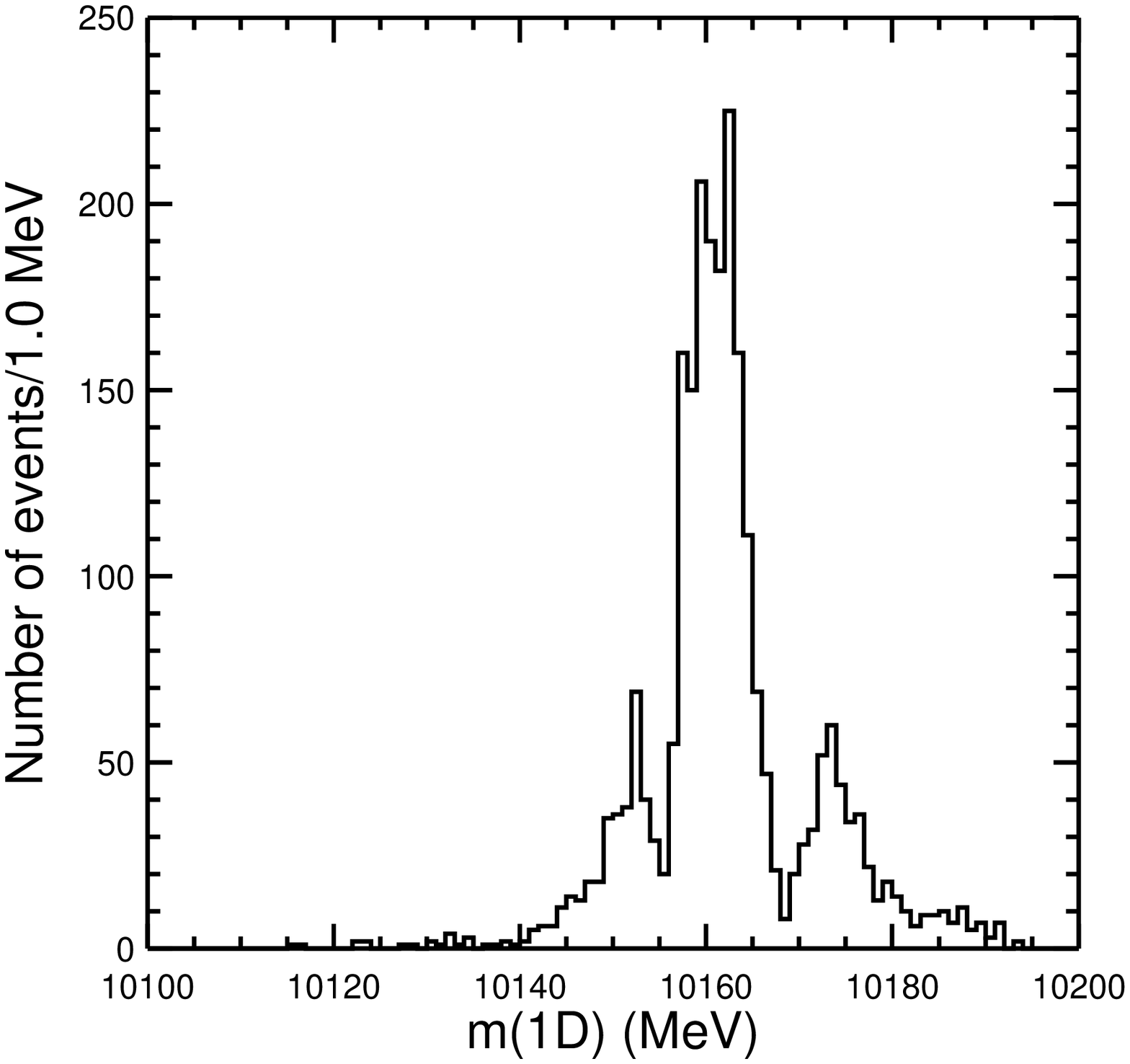}{Distribution of the
most likely mass assignment 
for the $\Y(1D_2)$ Monte Carlo events generated with a
mass of 10160 MeV/$c^2$. In addition to the narrow peak at the correct
mass, two satellite peaks at lower and higher mass are produced.
}{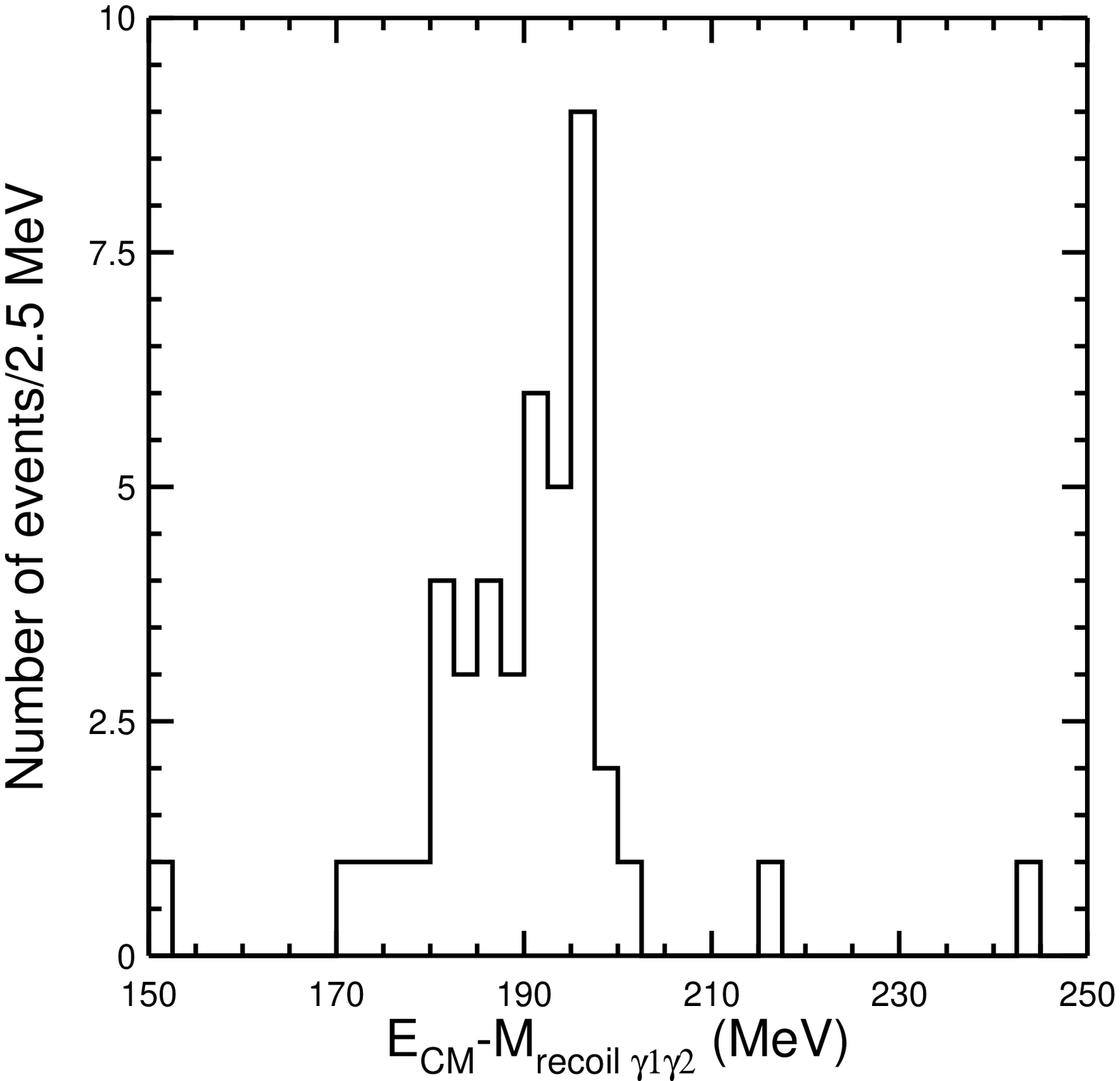}{Distribution of the difference between the 
center-of-mass energy and the recoil mass
against the two lowest energy photons
for the data.
}{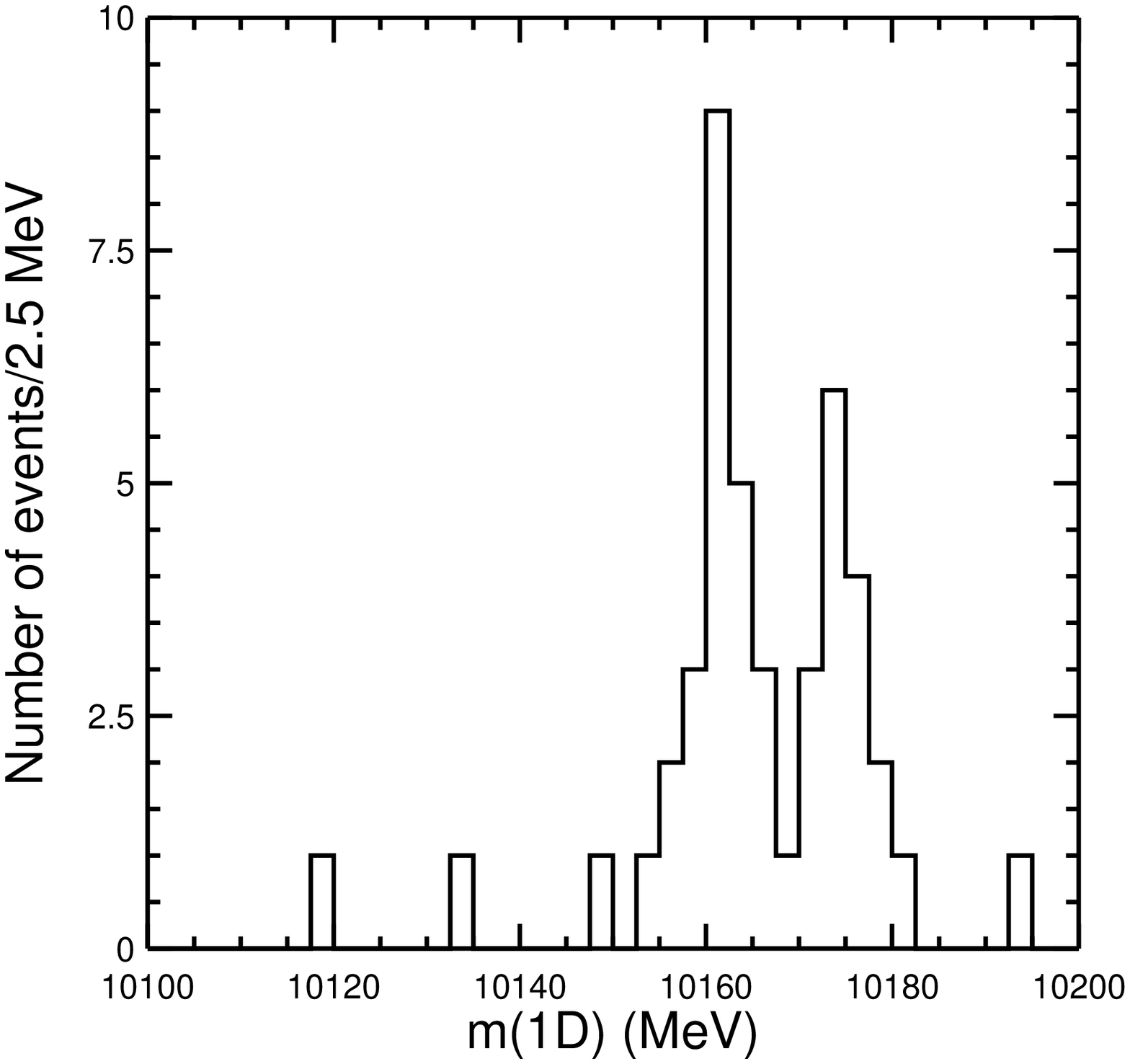}{Distribution of the 
most likely mass assignment 
for the $\Y(1D)$ candidate events in the data.
}

The distributions of $E_{CM}-M_{recoil\,\gamma_1\gamma_2}$
and the $\Y(1D)$ mass which minimizes $\chi^2_{1D}$ for each event
(called the most-likely mass)
are shown for the data in Figs.~\ref{fig:m3s1ddata}-\ref{fig:m1dfdata}
after the $\chi^2_{1D}<10$ cut.  
The background fraction for this sample is in the $9-14\%$ range. 
Two peaks are observed in the distribution of the most likely mass,
around 10160 and 10175 MeV/$c^2$.
The recoil mass distribution does not show a double-peak structure,
but the mass resolution is worse in this variable, as discussed
above. 

To analyze the most likely mass distribution, which has a better
mass resolution, we fit the data 
to either a two-peak or a one-peak (positioned either around
10160 or 10175 MeV/$c^2$) hypothesis. 
The background is assumed to
have a flat distribution. Since the backgrounds
are not significant, their parameterization is rather
unimportant. 
Because, in principle, up to three triplet $\Y(1D)$ states 
can contribute to our sample, we also tried to 
force a third peak into the fit function.
However, the fit either makes the amplitude of this
peak zero, with no effect on the amplitudes of the
other two peaks, or the third peak collapses into
one of the other two peaks.
The two-peak fit, displayed in Fig.~\ref{fig:m1dfit2},
gives the best confidence level  ($58\%$).
The amplitude (mass) of the first peak is
$27.8^{+6.8}_{-6.0}$ events ($10161.2\pm0.7$ MeV/$c^2$),
and of the second peak
$12.0^{+5.3}_{-4.6}$ events ($10174.2\pm1.3$ MeV/$c^2$).
The assumption that there is no peak around 10160 MeV/$c^2$ produces
a very low confidence level ($0.04\%$) and can be ruled out.
From the change of likelihood between these two fits,
the significance of the first peak is 6.8 standard
deviations. 
The hypothesis of just one peak around 10160 MeV/$c^2$ also produces 
a good confidence level of $43\%$ (see Fig.~\ref{fig:m1dfit1}),
with $38.6^{+6.8}_{-6.2}$ events in the peak 
($10162.0\pm0.5$ MeV/$c^2$ mass).
Furthermore, while the recoil mass distribution is consistent 
with the production of two states with the amplitudes and
masses as determined by the fit above 
(confidence level $36\%$), it prefers
to allocate most of the events to the lower mass state,
$37.7^{+8.5}_{-7.5}$, with only $2.9^{+5.8}_{-5.2}$ events
attributed to the higher mass state, when the amplitudes
are let free in the fit.
This fit is displayed in Fig.~\ref{fig:m3s1dfit2} 
and has a confidence level of $48\%$.
The single-peak interpretation of the recoil mass
distribution also has a high confidence level ($50\%$),
with $40.5^{+6.9}_{-6.3}$ events at a mass
of $10163.4\pm1.3$ MeV/$c^2$ (Fig.~\ref{fig:m3s1dfit1}).

In the region of the first peak in the most likely mass
distribution, $(77\pm9)$\%\
of the events have the $J_{2P}=1$, $J_{1P}=1$ combination as the
most likely. This is consistent with the Monte Carlo simulations for
both the $\Y(1D_2)$  ($78\%$) and  
$\Y(1D_1)$ ($73\%$) states.
Furthermore, we have very low efficiency for the $\Y(1D_3)$
state at this mass (Fig.~\ref{fig:effvsmass}). When we remove
the cuts suppressing the $\Y(2S)$ cascades, which are
responsible for the efficiency loss for the $\Y(1D_3)$ state, 
the amplitude of this
peak does not increase dramatically. Therefore,
this cannot be the $J_{1D}=3$ state.
The $\Y(1D_2)$ state is predicted to be produced with
a 6 times larger rate than the $\Y(1D_1)$ state \cite{GodfreyRosner}.
The Clebsch-Gordan coefficients play a big role in 
this ratio, thus the model dependence of this prediction is weak. 
Therefore, we take the $\Y(1D_2)$ state as the most likely interpretation
of the first peak. 
Taking an average over the different fits described above,
we measure mass of this state to be $10162.2\pm1.6$ MeV/$c^2$.
The error is dominated by the dependence on the fit approach ($\pm1.2$ MeV/$c^2$),
with smaller contributions from the statistical error ($\pm0.7$ MeV/$c^2$),
photon energy scale error ($\pm0.7$ MeV/$c^2$) and the uncertainty
on the $\Y(3S)$ mass ($\pm0.5$ MeV/$c^2$).
Predictions of the most successful potential models predict
the center-of-gravity of the $\Y(1D)$ triplet to be around
this mass \cite{GodfreyRosner}. Calculations of the fine 
splitting of the $\Y(1D)$ states predict the $J_{1D}=2$ state to be close
to the center-of-gravity mass \cite{GodfreyRosner}. 
Our mass measurement is consistent with these predictions.

Only 3 events have
$J_{2P}=2$, $J_{1P}=2$ as the most likely combination.
If the $J_{1D}=3$ state was produced with the amplitude
of 12 events, as suggested by the two-peak fit to the
distribution of the most likely mass,
we would expect 9.4 such events,   
including 1.5 events of feed-down from
the $\Y(1D_2)$ state at 10162 MeV/$c^2$.
As described previously,
the recoil mass distribution shows no evidence for
a state at 10174 MeV/$c^2$ either.
Thus, there is no strong evidence for the second state
in our data. 
More data, which will be accumulated by CLEO III 
at the $\Y(3S)$ in the future,
will help to clarify the interpretation of these results.

\global\def\1{m1dfit2}
\global\def\2{m1dfit1}
\global\def\3{m3s1dfit2}
\global\def\4{m3s1dfit1}
\quadruplex{htbp}{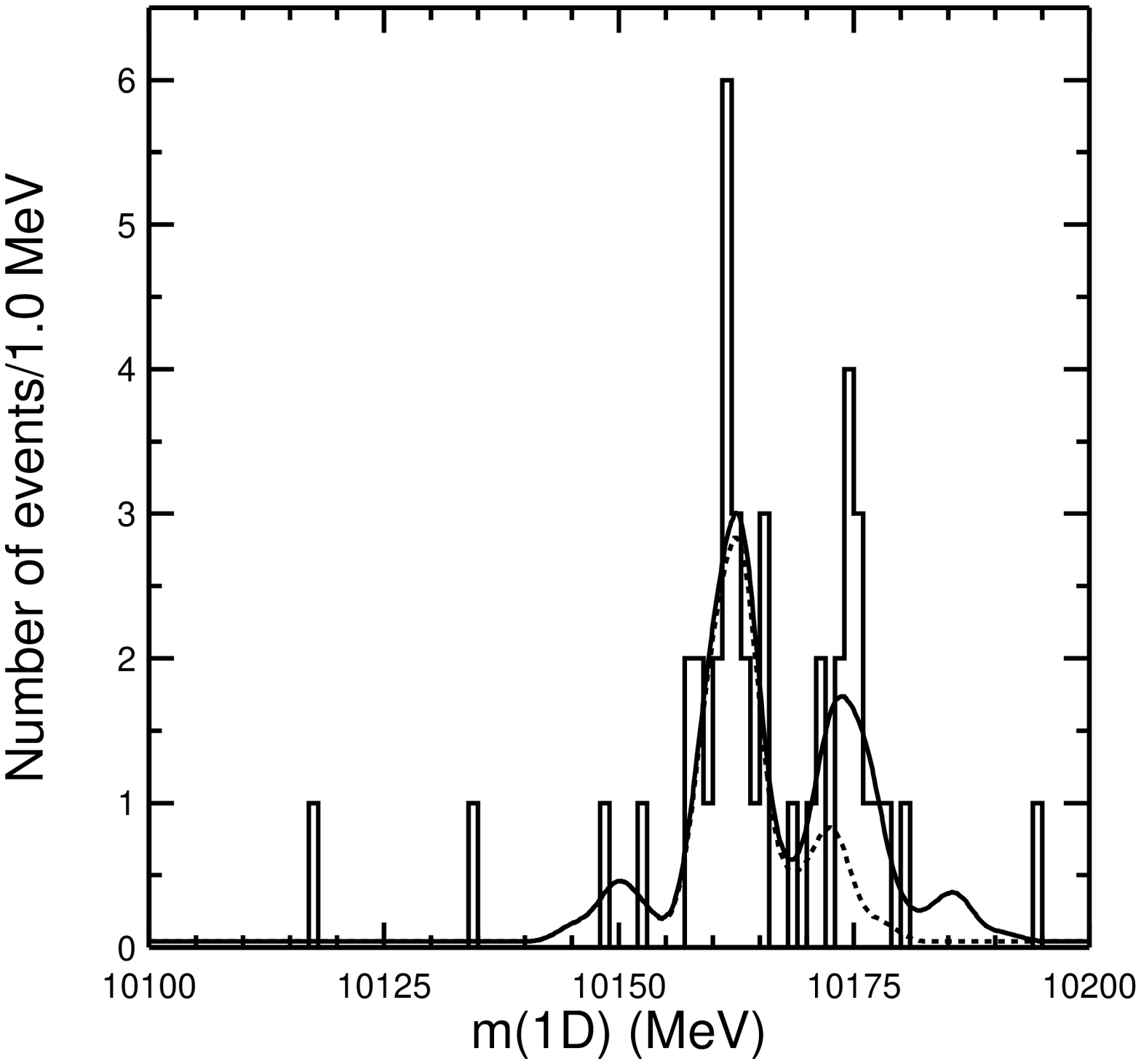}{
The fit (solid-line) of a two-peak structure to the observed
distribution of the most likely mass. 
The dashed-line indicates the fitted contribution of the
first peak plus the background.
The signal shape for each peak has a central peak, and
a smaller satellite peak on each side of the central peak.
}{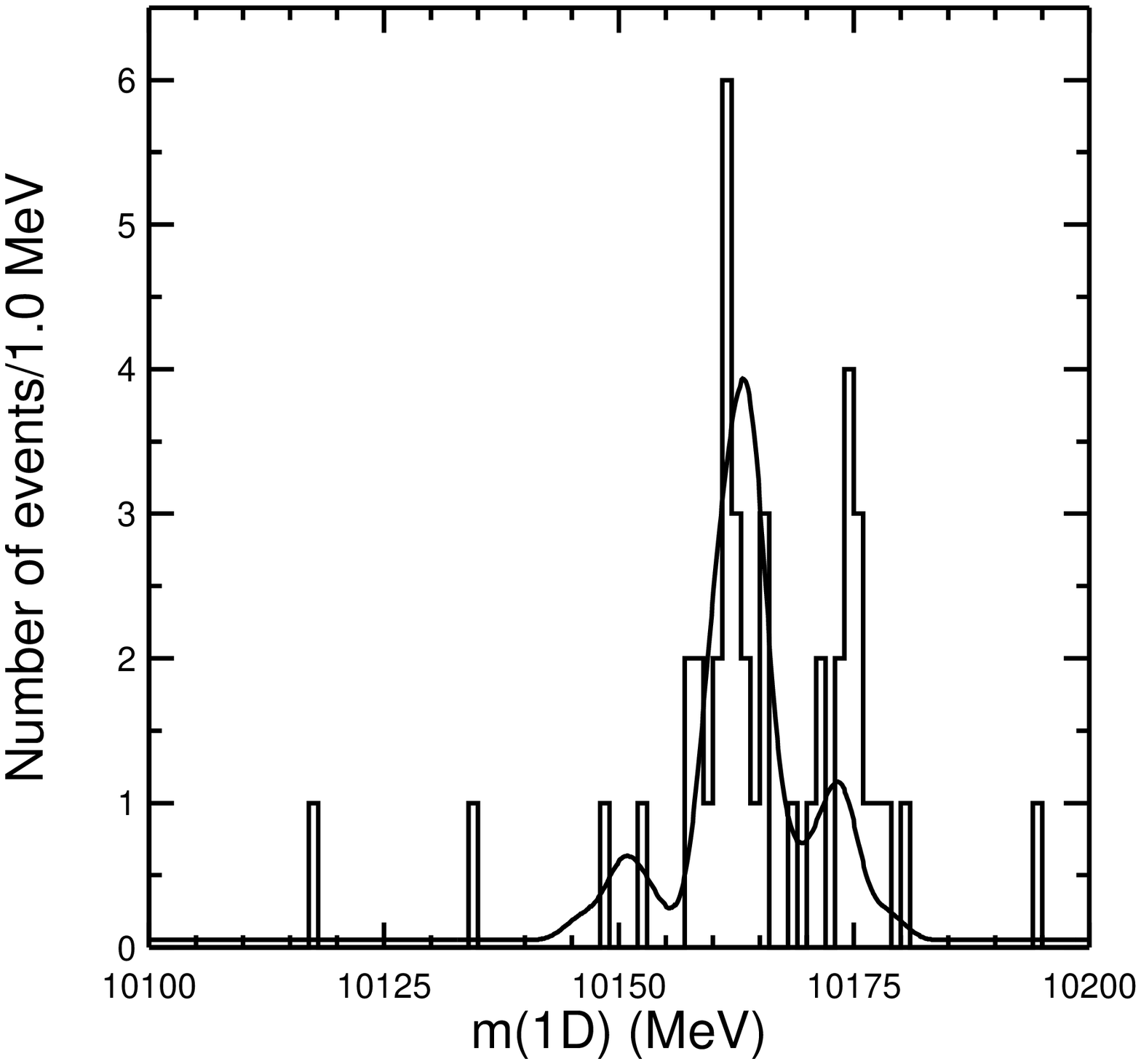}{
The fit (solid-line) of one-peak structure to the observed
distribution of the most likely mass. 
}{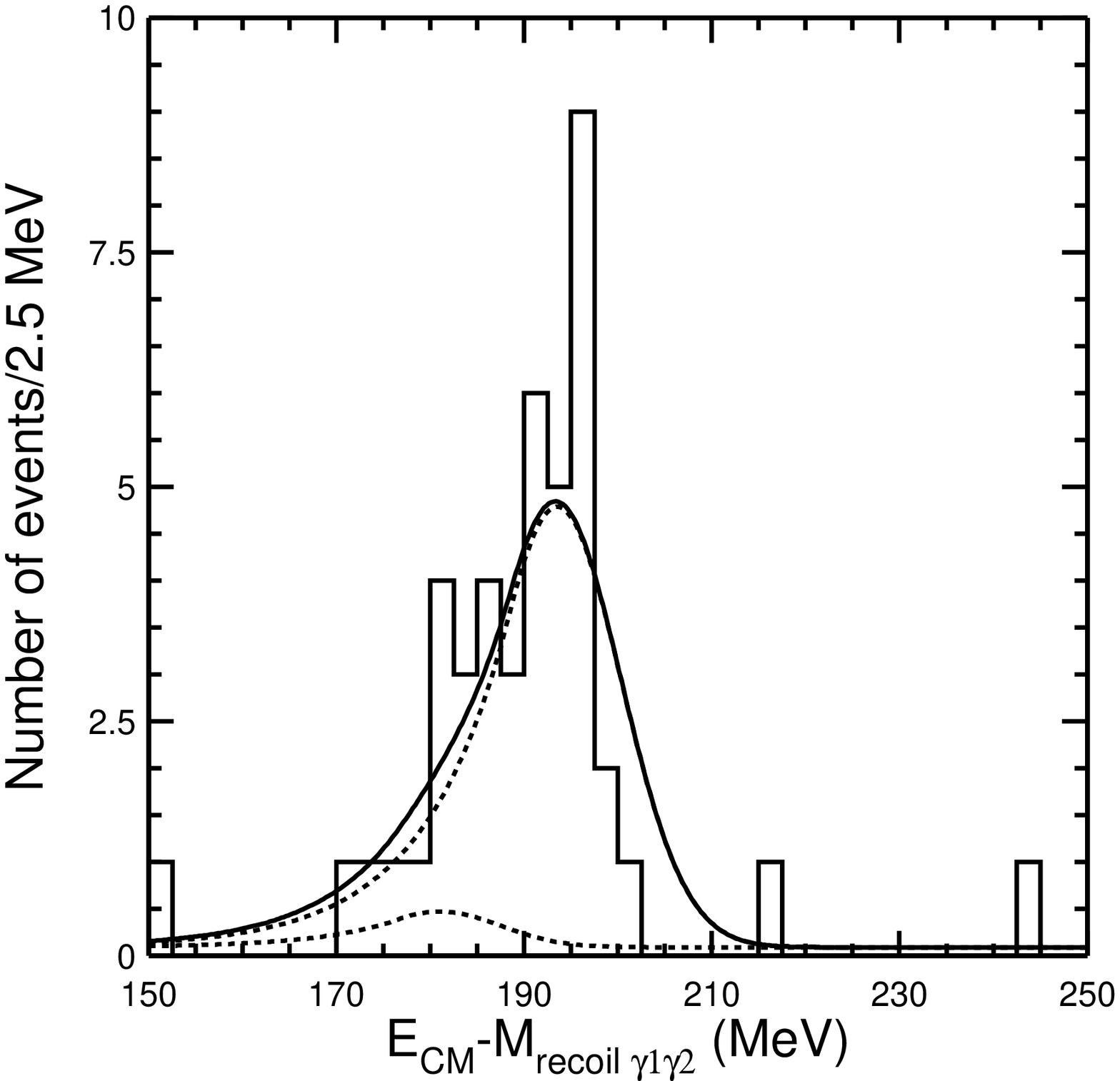}{
The fit (solid-line) of a two-peak structure to the
distribution of $E_{CM}-M_{recoil\,\gamma_1\gamma_2}$.
The peak positions were fixed to the masses obtained
by the fit to the most likely mass
(the higher mass state appears at a lower value of
$E_{CM}-M_{recoil\,\gamma_1\gamma_2}$).
The amplitudes of the peaks were allowed to float.
The dashed-lines indicate the fitted contributions of 
each peak separately.}{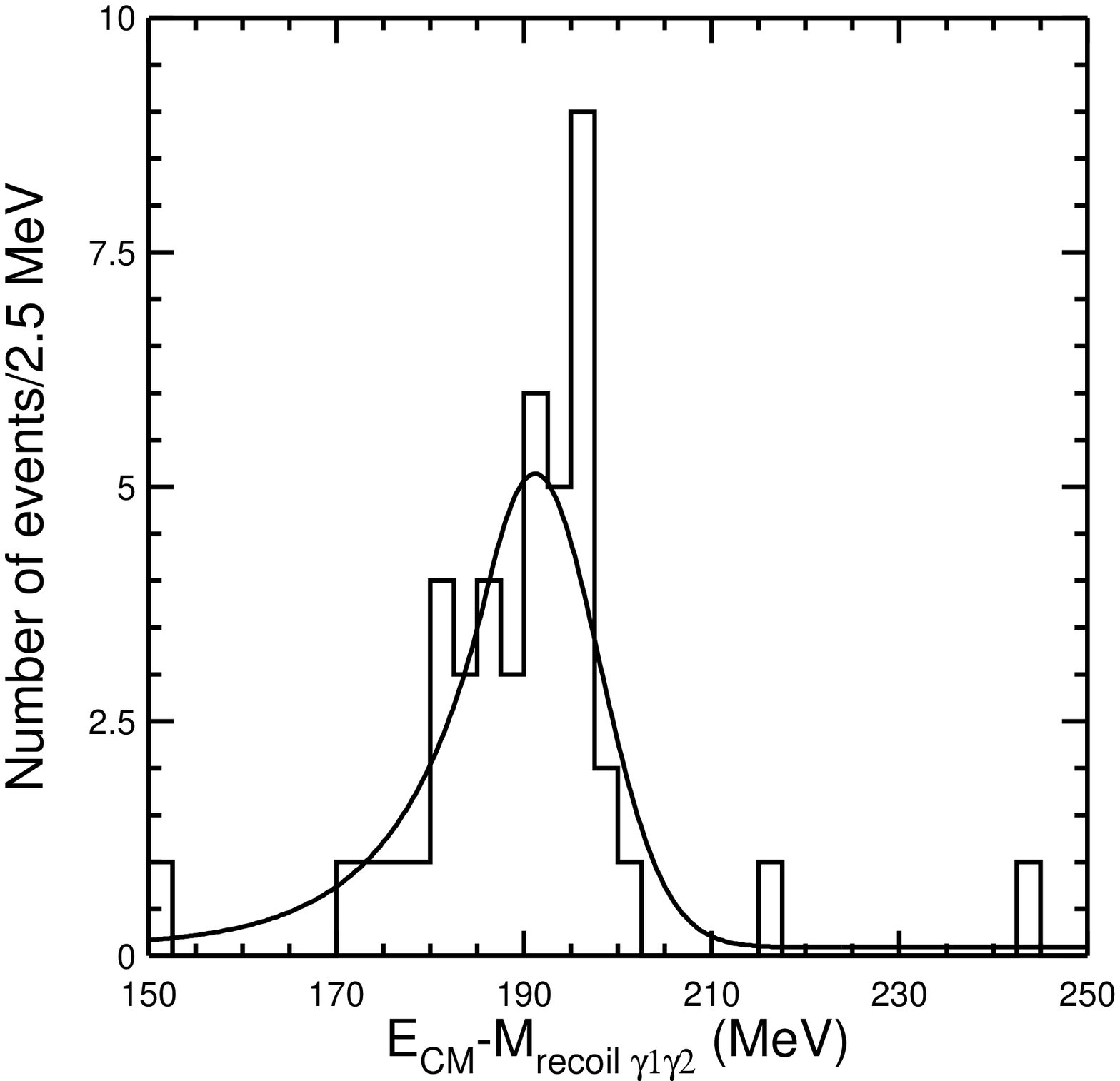}{
The fit (solid-line) of a one-peak structure to the observed
distribution of the difference between
the center-of-mass energy and the recoil mass against the
two lowest energy photons.
The amplitude and peak positions were free parameters in the fit.}

\section{Inclusive production rate for the $\Y(1D)$ states}

To calculate the signal efficiency in the fit to the $\chi^2_{1D}$
distribution (Fig.~\ref{fig:chidfit}), we weight the
$\Y(1D_2)$ efficiency for a state with a mass of $10160$ MeV/$c^2$ 
and the $\Y(1D_3)$ efficiency for a state with a mass of $10175$ MeV/$c^2$ 
in a ratio proportional to the observed peak amplitudes
in the two-peak fit to the observed most likely mass distribution.
We use the difference between this weighted efficiency
and the individual efficiencies for each state as an estimate
of the systematic error. This gives an efficiency of
$(13.2\pm1.0)\%$. Converting the number of $\Y(1D)$ events
determined by the fit to the $\chi^2_{1D}$ distribution to
a product  branching ratio, we obtain
$(3.3\pm0.6\pm0.5)\cdot10^{-5}$.
This is an inclusive rate that sums over all possible
$\Y(1D)$ states.
This rate is in good agreement with the $3.76\cdot 10^{-5}$
prediction by Godfrey and Rosner \cite{GodfreyRosner}.

\section{Summary}

In summary, we present evidence at the 
9.7 standard deviation significance for
production of $\Y(1D)$ states in four-photon cascades
from the $\Y(3S)$. 
The inclusively measured product branching ratio,
$\B(\Y(3S)\to\gamma\chi_b(2P))\cdot$
$\B(\chi_b(2P)\to\gamma\Y(1D))\cdot$
$\B(\Y(1D)\to\gamma\chi_b(1P))\cdot$
$\B(\chi_b(1P)\to\gamma\Y(1S))\cdot$
$\B(\Y(1S)\to l^+l^-)$
$=(3.3\pm0.6\pm0.5)\cdot10^{-5}$,
is in good agreement with the theoretical predictions by
Godfrey and Rosner \cite{GodfreyRosner}, where
$\B(\Y(1S)\to l^+l^-)$ above means the average branching ratio
over $e^+e^-$ and $\mu^+\mu^-$.
From the observed mass distribution we see a
6.8 standard deviation signal for a state with a mass
of $10162.2\pm1.6$ MeV/$c^2$. This is likely to be
the $J_{1D}=2$ state, though we cannot rule out the $J_{1D}=1$ hypothesis.
However, 
the $J_{1D}=1$ hypothesis is strongly disfavored by 
theoretical expectations.
Evidence for a second heavier state in our data is so
far inconclusive.
This is the first discovery of an $L=2$ quarkonium state
below the open flavor threshold.

We gratefully acknowledge the effort of the CESR staff in providing us with
excellent luminosity and running conditions.
We thank J.~L.~Rosner for the useful discussions concerning the $\Y(1D)$ states.
M. Selen thanks the PFF program of the NSF and the Research Corporation, 
and A.H. Mahmood thanks the Texas Advanced Research Program.
This work was supported by the National Science Foundation, and the
U.S. Department of Energy.

\end{document}